\documentstyle[prb,twocolumn,psfig,aps]{revtex}
\begin{document}
\draft

\twocolumn[\hsize\textwidth\columnwidth\hsize\csname @twocolumnfalse\endcsname

\title{Numerical Calculations of the $B_{1g}$ Raman Spectrum of the \\
Two-Dimensional Heisenberg Model}

\author{Anders W. Sandvik}
\address{Department of Physics, University of Illinois at Urbana-Champaign,
1110 West Green Street, Urbana, Illinois 61801}

\author{Sylvain Capponi and Didier Poilblanc}
\address{Laboratoire de Physique Quantique
and Unit\'e Mixte de Recherche CNRS 5626, 
Universit\'e Paul Sabatier, 31062 Toulouse, France}

\author{Elbio Dagotto}
\address{National High Magnetic Field Laboratory, Florida State
University, 1800 E. Paul Dirac Dr., Tallahassee, Florida 32306 }

\date{\today}

\maketitle

\begin{abstract}
The $B_{1g}$ Raman spectrum of the two-dimensional $S=1/2$ Heisenberg model 
is discussed within Loudon-Fleury theory at both zero and finite temperature. 
The exact $T=0$ spectrum for lattices with up to $6 \times 6$ sites is 
computed using Lancz\"os exact diagonalization. A quantum Monte Carlo (QMC)
method is used to calculate the corresponding imaginary-time correlation 
function and its first two derivatives for lattices with up to $16 \times 16$ 
spins. The imaginary-time data is continued to real frequency using the 
maximum-entropy method, as well as a fit based on spinwave theory.
The numerical results are compared with spinwave calculations for finite 
lattices. There is a surprisingly large change in the exact spectrum going 
from $4\times 4$ to $6 \times 6$ sites. In the former 
case there is a single dominant two-magnon peak at $\omega /J \approx 3.0$, 
whereas in the latter case there are two approximately equal-sized peaks at
$\omega /J \approx 2.7$ and $3.9$. This is in good qualitative agreement with 
the spinwave calculations including two-magnon processes on the same lattices. 
The spinwave results for larger lattices show how additional peaks emerge 
with increasing lattice size, and eventually develop into the well known 
two-magnon profile peaked at $\omega /J \approx 3.2$ and with weight
extending up to $\omega/J \approx 4.6$. Both the Lancz\"os and the QMC 
results indicate that the actual two-magnon profile is broader than 
the narrow peak obtained in spinwave theory, but the positions of the
maxima agree to within a few percent. The higher-order contributions 
present in the numerical results are merged with the two-magnon 
profile and extend up to frequencies $\omega /J \approx 7$. The first three 
frequency cumulants of the spectrum are in excellent agreement with results 
previously obtained from a series expansion around the Ising limit. Typical 
experimental $B_{1g}$ spectra for La$_{\rm 2}$CuO$_{\rm 4}$ are only slightly
broader than what we obtain here. The exchange constant extracted from the 
peak position is $J\approx 1400$K, in good agreement with values 
obtained from neutron scattering and NMR experiments. We discuss the 
implications of our present results for more sophisticated theories of 
Raman scattering suggested recently.
\end{abstract}

\vskip2mm


\vskip6mm

]

\section{Introduction}

The magnetic properties of the parent compounds of the high-T$_{\rm c}$ 
cuprate superconductors can be well accounted for by weakly coupled 
two-dimensional (2D) Heisenberg antiferromagnets.\cite{manousakis} Neglecting 
the weak interlayer coupling, the model is defined by the Hamiltonian
\begin{equation}
\hat H = J\sum\limits_{\langle i,j\rangle} {\bf S}_i \cdot {\bf S}_j,
\quad (J > 0),
\end{equation}
where ${\bf S}_i$ is a spin-$1/2$ operator at site $i$ on a square lattice and
${\langle i,j\rangle}$ denotes a pair of nearest-neighbor sites. The most well
studied among the antiferromagnetic layered cuprates is La$_2$CuO$_4$,
\cite{neutrons1,neutrons2,neutrons3,imai,matsu} with a N\'eel ordering
temperature $T_N \approx 300$ K. For $T > T_N$, the temperature dependence 
of the spin correlation length, measured using neutron scattering,
\cite{neutrons1} is in good agreement with that of a 
single-layer Heisenberg model with $J \approx 1500$ K.\cite{csh,ding} 
The spinwave spectrum of the Heisenberg model is well reproduced over the 
entire Brilloin zone.\cite{neutrons2,makivic} The NMR relaxation rates 
$1/T_1$ and $1/T_{2G}$, which probe the low-frequency spin dynamics, also 
show remarkable agreement between experiment\cite{imai,matsu} and theory.
\cite{csh,qc,sokol,qmcrates} 

In contrast to these success stories, the experimental Raman spectrum
\cite{ramanexp1,knoll,blumberg} 
shows significant deviations from calculations for 
the 2D Heisenberg model.\cite{parkinson,canali,roger,diag} Within this 
description of the CuO$_{\rm 2}$ layers, the standard theory of Raman 
scattering is based on the Loudon-Fleury (LF) coupling \cite{fleury} between 
the light and the spin system. The coupling is obtained in second order 
perturbation theory with virtual states containing one doubly occupied site, 
and is given by \cite{fleury,shastry}
\begin{equation}
\hat H_{LF} = \sum\limits_{\langle i,j\rangle} 
({\bf E}_{\rm in} \cdot {\bf \sigma}_{ij})
({\bf E}_{\rm out} \cdot {\bf \sigma}_{ij})
{\bf S}_i \cdot {\bf S}_j.
\end{equation}
Here ${\bf E}_{\rm in}$ and ${\bf E}_{\rm out}$ are the polarization vectors
of the incoming and scattered light and ${\bf \sigma}_{ij}$
is the unit vector connecting sites $i$ and $j$. In terms of the eigenstates
$\lbrace |n \rangle\rbrace$ of the Heisenberg model, the frequency dependence 
of the scattering intensity at inverse temperature $\beta$ is given by 
Fermi's Golden Rule: 
\begin{eqnarray}
I(\omega) = & & {1\over Z} \sum\limits_{m} 
{\rm e}^{-\beta E_m} \times \nonumber \\
& & \sum\limits_n |\langle n | \hat H_{LF} | m \rangle |^2 
\delta (\omega - [E_n - E_m]).
\label{fermi}
\end{eqnarray}
Most theoretical work has focused on the scattering in the $B_{1g}$ symmetry
channel. This corresponds to ${\bf E}_{\rm in}$ along a diagonal of the 
square lattice, and ${\bf E}_{\rm out}$ perpendicular to ${\bf E}_{\rm in}$. 
The $B_{1g}$ coupling can thus be written as
\begin{equation}
\hat H_{LF} = \sum\limits_{~\langle i,j\rangle_x} {\bf S}_i \cdot {\bf S}_j
- \sum\limits_{~\langle i,j\rangle_y} {\bf S}_i \cdot {\bf S}_j,
\label{b1g}
\end{equation}
where ${\langle i,j\rangle_x}$ and ${\langle i,j\rangle_y}$ denote links in 
the $x$ and $y$ directions, respectively. 

For La$_2$CuO$_4$, the $B_{1g}$ spectrum has a broad asymmetric peak at 
$\omega \approx 3J$ with a tail extending to $\omega \approx 7-8J$.
\cite{ramanexp1} In some cases there is a shoulder-like structure at 
$\omega \approx 4J$. Within spinwave theory the $B_{1g}$ scattering is 
dominated by two-magnon excitations. The two-magnon profile is 
peaked around $\omega \approx 3J$ in good agreement with the experiments.
\cite{parkinson,canali,roger,diag} However, the large width of the 
experimental spectrum has not been reproduced within spinwave theory. The
first three frequency cumulants of the spectrum have also been calculated 
using  a series expansion around the Ising limit,\cite{singh} and are in 
reasonable agreement with the experimental values. The moments obtained
in spinwave theory to order $1/S$ (two magnon excitations only) are in poor 
agreement with the series results which in principle include multi-magnon 
contributions to all orders. Unfortunately, the full frequency dependence is 
not accessible with the series expansion method. Exact diagonalization 
has been used to compute the exact LF Raman profile for small lattices. 
\cite{diag} For the $4\times 4$ system there is a single dominant two-magnon 
peak at $\omega /J =2.98$. Weight present at $\omega \approx 5J$ has been 
attributed to four-magnon processes, but its relative strength 
is much smaller than the weight found experimentally in this frequency 
region. The tail at higher frequencies is absent. Despite this, the 
first three frequency cumulants are in approximate agreement 
with both experiments and the series expansion results.

Canali and Girvin\cite{canali} carried out a spin-wave expansion including 
also four-magnon excitations (which enter in order $S^{-2}$). The narrow width 
of the two-magnon peak was found to be stable with respect to inclusion of the
higher-order processes. The relative contribution from four-magnon states 
in this calculation is less than $3\%$. The high energy of the four-magnon
weight nevertheless leads to first and second cumulants that are much closer 
to those obtained in the series expansion and exact diagonalization studies. 
The spinwave result for the third cumulant is, however, significantly larger 
than the series expansion value. It was argued that this is due to 
interactions between four magnons that were neglected in the $1/S^2$ 
calculation, and that the relative four-magnon contribution must be 
$\approx 10\%$ in order to reproduce the first three moments.\cite{canali} 
The conclusion that there is less high-energy weight than in typical 
experimental spectra then still remains. However, the apparent inability of a 
very sophisticated spinwave calculation to fully capture the four-magnon 
processes raises some concerns about this approach for calculating the actual
line shape. Furthermore, Chubukov and Frenkel have recently questioned the 
stage at which the spin $S$ was set to $1/2$ in the previous spinwave 
calculations. They kept $S$ large and carried out an expansion of the profile 
around its peak position before expanding in $1/S$ and evaluating the result 
at $S=1/2$. The two-magnon profile obtained this way has a width almost three
times larger than the ``standard'' one, and the second frequency cumulant is 
therefore in better agreement with the series result. However, the agreement 
with the first cumulant is actually worse, due to the significantly larger 
low-frequency weight. Hence, this result also has to be viewed with some 
caution.

Experimentally, significant scattering is also observed in the $A_{1g}$ 
channel.\cite{ramanexp1} With the standard LF coupling inelastic scattering 
in this symmetry is not possible for the Heisenberg model with 
only nearest-neighbor interactions, since the $x$ and $y$ terms in
Eq.~(\ref{b1g}) are added in this case and $\hat H_{LF}$ then commutes
with the Hamiltonian. Adding a next-nearest-neighbor term to the LF coupling 
leads to $A_{1g}$ scattering, but has no effect in the $B_{1g}$ channel.
\cite{singh} Based on the frequency moments obtained in the series expansion, 
this was argued to be a mechanism that could explain both the $B_{1g}$ and 
$A_{1g}$ spectra. However, the narrow $B_{1g}$ line-shape obtained in other 
calculations remains an 
unresolved issue for this scenario. Including a next-nearest-neighbor term 
$\sim J_2$ in the Hamiltonian would also lead to $A_{1g}$ scattering, and 
probably a broadened $B_{1g}$ spectrum. The relatively large $A_{1g}$ 
intensity seen experimentally would then likely require a larger $J_2$ than 
allowed by other experiments, but detailed 
calculations have not been carried out within this model. Other interactions,
such as the so called four-spin cyclical exchange,\cite{roger,honda} have 
also been suggested to account for the differences between theory and 
experiment. Analytical calculations as detailed as those for the standard 
LF theory (only nearest-neighbor interactions in both the Hamiltonian and 
the Raman operator) have also not yet been carried out within these
theories. 

It is clear that LF theory is not sufficient to capture all 
aspects of Raman scattering in layered cuprates. For example, resonant
scattering (occurring when the frequency of the incoming light is comparable 
to the charge-transfer gap)  can of course not occur within the Heisenberg
model.\cite{chubukov}  However, focusing on nonresonant $B_{1g}$ scattering, 
it is not yet clear what the actual line shape is within LF theory. As
discussed above, the two key questions of the width of the dominant two-magnon
profile and the relative weight of the higher-order contributions
above the two-magnon cut-off remain incompletely answered within spinwave 
theory. The exact diagonalization studies carried out so far are limited to 
lattices too small for reliable quantitative extrapolation to the 
thermodynamic limit. There is hence a definite need for accurate 
non-perturbative numerical calculations for larger lattices. Conclusive 
results would provide a more solid basis for estimating effects not 
included in the LF-Heisenberg model. It should also be noted that LF theory 
is the standard framework in which Raman scattering has been interpreted
also in several other low-dimensional antiferromagnetic $S=1/2$ 
systems.\cite{prelsingh,gros} A satisfactory resolution of 
the 2D Heisenberg case would therefore be of more general interest as well. 
Finally, knowing the exact $B_{1g}$ Raman spectrum should help to shed light 
on the applicability of spinwave theory for calculations of dynamic processes
involving excitations of more than one magnon.

Here we report new exact diagonalization results for systems with up to 
$6 \times 6$ spins, which is the largest currently accessible with this 
method. We also obtain approximate spectra for up to $16 \times 16$ spins by 
Maximum-Entropy (Max-Ent) analytic continuation of quantum Monte Carlo (QMC) 
data. Using the stochastic series expansion QMC technique 
\cite{qmc1,qmc2,sandvik} we have calculated the imaginary-time LF correlation 
function and its first 
two derivatives at temperatures low enough to give ground state results. 
The derivatives are used as supplementary information in the analytic 
continuation. We also consider a more phenomenological approach of fitting 
the two-magnon profile obtained within spinwave theory to the imaginary-time 
data, adding a Gaussian at higher frequency to model the higher-order 
contributions. In order to study the effects of temperature, we apply the 
QMC + Max-Ent methods also at non-zero temperatures.

We find that the $B_{1g}$ Raman spectrum of the $6 \times 6$ lattice has 
two dominant peaks at $\omega /J \approx 2.7$ and $3.9$, in sharp contrast to 
the single dominant peak at $\omega/J=2.98$ previously found for the 
$4 \times 4$ lattice.
These results are qualitatively reproduced within spin wave theory including 
only two-magnon excitations and magnon-magnon interactions (treated within 
an RPA scheme). Spinwave results resembling the infinite-size two-magnon 
profile are seen only for much larger lattices. Using the QMC data, the first
three frequency cumulants in the thermodynamic limit can be reliably
estimated. They are in excellent agreement with the previous series expansion
results.\cite{singh} Both the Lancz\"os and the QMC results indicate that the 
dominant peak at $\omega \approx 3.3J$ is slightly broader than the standard 
two-magnon profile, but not as broad as the the one recently obtained by 
Chubukov and Frenkel.\cite{chubukov}  We find no evidence for a gap between 
the two-magnon profile and the higher-order contributions. They appear to be 
completely merged together and extend up to $\omega \approx 7J$.

Finite-temperature results for $T/J \alt 0.25$ are very similar
to the ground state results. For higher temperatures there is a significant 
growth of the low-frequency spectral weight, as also found in a previous exact 
diagonalization study of a $4 \times 4$ lattice.\cite{bacci} We find that 
the temperature at which this effect becomes significant decreases with 
increasing system size, due to the large finite-size gaps present 
in the smaller systems. 

Over-all, our results are in closer agreement with experiments than previous 
exact diagonalization and spinwave calculations, but we conclude that the 
Heisenberg-LF spectrum is nevertheless not quite as broad as typical 
experimental spectra. We argue that our results give more credibility to 
proposed broadening mechanisms involving phonons.
\cite{knoll,weber,nori,manning,saenger} 

The rest of the paper is organized as follows. In Sec.~II we review various 
spinwave calculations of the $B_{1g}$ Raman profile. We also present results
for small lattices, which are compared with exact diagonalization spectra in
Sec~III. Our $T=0$ results from QMC simulation and numerical analytic 
continuation are discussed in Sec.~IV. Effects of finite temperature are 
considered in Sec.~V. In Sec.~VI we summarize and discuss our results and 
implications for mechanisms proposed to lead to a broadening of the $B_{1g}$ 
profile. In an Appendix we present some technical details of the QMC 
calculation of the imaginary-time correlation function and its derivatives.

\section{Spinwave Theory}

The Raman spectrum can be easily computed in the spinwave approximation.
Various improvements of the linear spinwave calculation can be also applied 
to the Raman scattering amplitude. For example, the residual interactions 
between spin waves, which play a crucial role in the Raman excited states, 
can be included at the RPA level. Here we discuss linear spin wave theory 
and the effect of magnon-magnon interactions on the $B_{1g}$ spectrum. 
The primary purpose of the calculations discussed here is to qualitatively 
understand the effects of finite size, which will be important for 
interpreting the numerical results presented in the following sections. 
More sophisticated spinwave calculations including the quantum fluctuations 
of the ground state, as well as final states with four magnons,  have been 
carried out before, as discussed in the Introduction. Here we note some 
problems with the analytical calculations of the Raman profile which 
motivate our renewed efforts to obtain accurate non-perturbative numerical 
results.

In the antiferromagnetic ground state we are considering, a bosonic 
representation of the spin operators can be introduced on each sublattice 
by using the usual Dyson-Maleev transformation~\cite{dyson,maleev}. 
On sublattice $A$, the transformation reads
\begin{mathletters}
\begin{eqnarray}
S_i^z & = & S - a_i^\dagger a_i \\
S_i^+ & = &  \sqrt{2S}\Bigl( 1-\frac{a_i^\dagger a_i}{2S}\Bigr) a_i \\
S_i^- & = & \sqrt{2S}a_i^\dagger ,
\end{eqnarray}
\end{mathletters}
where $a_i^\dagger$  creates a boson, i.e., a spin-1 magnon, at site $i$.
Similarly, on sublattice $B$, 
\begin{mathletters}
\begin{eqnarray}
S_j^z & = & -S + b_j^\dagger b_j \\
S_j^+ & = & \sqrt{2S} b_j^\dagger \Bigl( 1-\frac{b_j^\dagger b_j}{2S}\Bigr) \\
S_j^- & = & \sqrt{2S}b_j .
\end{eqnarray}
\end{mathletters}

As usual, as long as small fluctuations around the N\'eel ordered phase are 
considered, one keeps only quadratic terms in the Hamiltonian (which are the 
dominant terms in a $1/S$ expansion). Therefore, in this approximation, the 
Hamiltonian can be diagonalized by a Bogoliubov transformation in terms of 
spinwave excitations (or magnons);
\begin{mathletters}
\begin{eqnarray}
\alpha_{\bf k}&=&u_{\bf k} a_{\bf k}+v_{\bf k} b_{\bf k}^\dagger\\
\beta_{\bf k}&=&u_{\bf k} b_{\bf k}+v_{\bf k} a_{\bf k}^\dagger \, ,
\end{eqnarray}
where the coefficients ($>0$) are given by,
\begin{eqnarray}
u_{\bf k}^2&=&\frac{1}{2}\Bigl(\frac{1}{\strut \sqrt{1-\gamma_{\bf k}^2}}+1 
\Bigr)\\
v_{\bf k}^2&=&\frac{1}{2}\Bigl(\frac{1}{\strut \sqrt{1-\gamma_{\bf k}^2}}-1 
\Bigr)  \, .
\end{eqnarray}
\end{mathletters}
We have defined $ \gamma_{\bf k}=\sum_{\bf \delta} e^{i {\bf k} \cdot
{\bf \delta}} / Z$ as a sum over the $Z$ nearest neighbours of the site
at the origin. In our case (square lattice), 
$\gamma_{\bf k}=(\cos k_x + \cos k_y)/2$. The resulting well known linear
spinwave Hamiltonian reads, 
\begin{equation}
H_{SW}=E_0 + \sum_{\bf k} \omega_{\bf k} (\alpha_{\bf k}^{\dagger} \alpha_{\bf
k}^{\phantom{\dagger}} + \beta_{\bf k}^{\dagger} \beta_{\bf k}^{\phantom{\dagger}}) ,
\label{HSW}
\end{equation}
with the dispersion relation $\omega_{\bf k}=JSZ\sqrt{1-\gamma_{\bf k}^2}$.
Due to the decomposition into two sublattices, the reciprocal space is reduced 
to the magnetic brillouin zone (MBZ).

The LF operator can be easily expressed as a quadratic form in terms of 
the spin wave operators. At zero temperature the ground state contains no 
bosons if magnon-magnon interactions are neglected so that, in a first 
approximation, one only keeps constant terms or terms involving 
the creation of magnons, giving
\begin{eqnarray}
H_{LF} & = &-\alpha NS^2  {\bf E}_{\rm out}\cdot {\bf E}_{\rm in}\\
& + & 2\alpha S\sum_{\bf k} \Bigl({E}_{\rm out}^x 
{E}_{\rm in}^x[\cos k_x(u_{\bf k}^2+v_{\bf k}^2)-
2u_{\bf k} v_{\bf k}]\\
&+& {E}_{\rm out}^y {E}_{\rm in}^y[\cos k_y
(u_{\bf k}^2+v_{\bf k}^2)-2u_{\bf k}
v_{\bf k}]\Bigr)\,\,\alpha_{\bf k}^\dagger \beta_{\bf k}^\dagger .
\end{eqnarray}
In the $B_{1g}$ geometry, the matrix element is thus
\begin{equation}
\langle f | H_R | i \rangle = \frac{\cos k_x - \cos
k_y}{\sqrt{1-\gamma_{\bf k}^2}} .
\end{equation}
The Raman intensity obtained from Fermi's Golden Rule, Eq.~(\ref{fermi}),
is then
\begin{equation}
I(\omega) \propto \sum_{\bf k} \frac{(\cos k_x - \cos
k_y)^2}{1-\gamma_{\bf k}^2} \delta(\omega-2 \Omega_{\bf k}),
\label{inonint}
\end{equation}
where $\Omega_{\bf k}=4 J S \sqrt{1-\gamma_{\bf k}^2}$ is the frequency of
the magnon. This expression exhibits a divergence at $\omega=8 J S$ since 
the density of states diverges at the boundary of the MBZ.

It is well known that this result is strongly modified when one
takes into account the magnon-magnon interactions in the final state.
\cite{elliott,parkinson,canali,chubukov} In order $1/S$, the Dyson-Maleev 
transformation generates in the Hamiltonian quartic terms in the bosons 
operators. One way of treating this 
interaction is to keep the term 
$\alpha^\dagger_{\bf k } \beta^\dagger_{\bf k } 
\beta_{-\bf p }^{\phantom{\dagger}}
\alpha_{\bf p }^{\phantom{\dagger}}$~\cite{parkinson,canali,chubukov} 
which is responsible for multiple scattering of two magnons in the vacuum.
This part gives the dominant contribution to the magnon-magnon scattering 
coming from the region near the MBZ boundary where the density of
states diverges. Further simplification results from the vanishing
of $\gamma_{\bf k}$ at the MBZ boundary so that it is reasonable to
replace $\gamma_{\bf k}$ by zero for all $\bf k$. This leads to an 
effective interaction of the form, \cite{chubukov}
\begin{equation}
H_{\rm int}=-\frac{4J}{N}\sum_{\bf k}\sum_{\bf p}\gamma_{\bf k-p} 
\alpha^\dagger_{\bf k }
\beta^\dagger_{\bf k } \beta_{-\bf p }^{\phantom{\dagger}}
\alpha_{\bf p }^{\phantom{\dagger}} .
\end{equation}
Following Refs.~\onlinecite{chubukov,canali} and expressing 
$\gamma_{\bf k-p}$ as a function of its symmetric terms
$\gamma_{\bf k-p}=\gamma^{c+}_{\bf k} \gamma^{c+}_{\bf p}
+\gamma^{c-}_{\bf k} \gamma^{c-}_{\bf p}
+ \gamma^{s+}_{\bf k} \gamma^{s+}_{\bf p} +\gamma^{s-}_{\bf k}
\gamma^{s-}_{\bf p}$  
\begin{eqnarray}
\gamma^{c\pm}_{\bf k}&=&(\cos k_x \pm \cos k_y)/2\\
\gamma^{s\pm}_{\bf k}&=&(\sin k_x \pm \sin k_y)/2  \, ,
\end{eqnarray}
it can be shown that multiple diffusion RPA series only contains 
terms involving  $\gamma^{c-}_{\bf p}$ factors. The final RPA expression 
for the Raman intensity is given by,\cite{chubukov}
\begin{equation}
I(\omega) \propto {\rm Im} \left\lbrace \frac{R(\omega )}{1+ R(\omega)/4S}
\right\rbrace
\label{iomega}
\end{equation}
with
\begin{equation}
R(\omega )=\frac{8JS}{N}\sum_{\bf k} \frac{(\cos k_x - \cos
k_y)^2}{\omega-2\Omega_{\bf k} +i\varepsilon} .
\label{specs1}
\end{equation}
In the thermodynamic limit, Eq.~(\ref{specs1}) for $S=1/2$ leads to a narrow 
two-magnon peak around $\omega=2.78 J$ which extends up to $\omega/J = 4$. 
In order to be able to directly 
compare spinwave results with exact spectra obtained with the Lancz\"os 
diagonalization method and approximate results of QMC and Max-Ent analytic 
continuation (presented in the following two sections), we have also evaluated 
(\ref{iomega}) for small lattices. Results for $L\times L$ clusters with 
$L=4,6,8$ and $10$ are shown in Fig.~\ref{figspinwave}, along with 
the corresponding non-interacting form (\ref{inonint}). It is clear that the 
continuous absorption band for infinite size is obtained from the accumulation
of a series of peaks for increasing cluster sizes. However, the number of 
peaks is still very small even for a lattice of 100 sites.

In the above calculation the magnon-magnon interactions have only been 
included in the final two-magnon state. The main effect of interactions
in the ground state is to renormalize the spinwave velocity; $c \to Z_cc$, 
where to order $1/S$  the renormalization factor  $Z_c = 1.158$ and arises 
solely from the normal ordering of the quartic magnon terms.
\cite{oguchi}  Hence, in a 
phenomenological way, the $1/S$ corrections can be taken into account by 
shifting the energy scale by a factor $Z_c$, leading to a $B_{1g}$ profile
peaked at $\omega/J = 3.22$. However, Canali and Girvin have shown that
the renormalization in fact enters in a non-trivial way in the expression
for the Raman profile, and hence there are other effects as well.\cite{canali} 
Nevertheless, the end result for $S=1/2$ does not differ much from 
Eq.~(\ref{specs1}) with $Z_c=1.158$, as shown in Fig.~\ref{figspinwave}.

Chubukov and Frenkel have recently raised questions about the stage at which 
one should set $S=1/2$ in the spinwave calculation. They argued that one 
should first expand the large-S expression for the profile (\ref{iomega}) 
around its peak position, and only after that set $S=1/2$.\cite{chubukov} 
The position of 
the maximum then remains approximately the same, but the profile is 
considerably broader, as also shown in Fig.~\ref{figprofiles}. The better 
agreement with the frequency moments calculated by Singh {\it et al.} 
\cite{singh} was 
used as support in favor of the broader peak. However, it should be noted
that the agreement with the first cumulant is actually significantly worse, 
due to the much slower decay of the weight on the low-frequency side of the 
peak. In fact, as seen in Fig.~\ref{figprofiles}, the Chubukov-Frenkel 
profile extended towards lower frequencies in a way similar to the
experimental spectrum, but the high-frequency tail is still of course missing.
The high-frequency scattering was argued to be dominated by resonant 
scattering in typical experimental situations, and would hence not be 
explainable by LF theory.\cite{chubukov}

To higher orders in $1/S$, states enter in which $4$, $6$, e.t.c.,
magnons are excited (the true ground state is a linear combination of
states containing any even number of magnons, and the Raman operator can
create or annihilate one or two pairs of magnons). Canali and Girvin included 
four-magnon excitations (order $1/S^2$) but neglected interactions involving 
more than two magnons.\cite{canali}
The two-magnon profile obtained this way is very 
similar to the Canali-Girvin $1/S$ result discussed above. The relative 
contribution from four-magnon processes is less than $3$\%, but is likely 
strongly affected by the neglected interactions.\cite{canali} In fact, although
the small four-magnon contribution is sufficient (because of its rather 
high energy) to change the first and second frequency cumulants to values 
in close agreement with the series expansion results, the third cumulant
remains far off. It was argued that this inconsistency is due to the neglected
interactions among four magnons (which may even lead to bound states), and 
that such interactions would bring the four-magnon peak position down in 
frequency.\cite{canali} The relative four-magnon weight would then have to 
increase to $\approx 10$\% in order to satisfy the first three moments. In 
our opinion, this rough estimate indicates that the approximations made in 
the $1/S^2$ calculation may in fact be serious. In particular, it is not 
clear that the two-magnon and four-magnon contributions will be well 
separated from each other if the four-magnon weight moves down and increases 
by a factor of 3 or more. This, in turn, may lead to considerably stronger 
interference effects that may cause changes also to the upper edge of the 
two-magnon profile (which then no longer would arise from two-magnon 
excitations only).

There are hence two major concerns with the spinwave calculations that
have to be addressed: 1) The stage at which $S$ is set to $1/2$, leading to 
two very different two-magnon line shapes. 2) The contributions from processes
including four or more magnons, which are very difficult to capture completely
within spinwave theory. In this situation it is clearly useful to consider 
non-perturbative numerical methods. We discuss two complementary approaches
in the next two sections.

\section{Exact Diagonalization}

In this section, we compute the exact Raman spectrum on clusters with up to
$N=36$ sites by use of the Lancz\"os diagonalization algorithm. In this 
approach, the Raman spectrum is obtained from a continued fraction,
\begin{equation}
I(\omega)=-\frac{1}{\pi} {\rm Im}\left\lbrace 
\langle 0|\hat H_{LF}^\dagger 
\frac{1}{\omega+E_0+i\varepsilon-\hat H}\hat H_{LF}|0\rangle \right\rbrace,
\end{equation}
where $|0\rangle$ is the ground state of energy $E_0$ which can be 
easily calculated with the Lancz\"os method, and $\varepsilon$ is a small 
imaginary part added to give a finite damping of the $\delta$-functions.

Results for several square and tilted lattices are shown in 
Fig.~\ref{figlanczos}, along with the RPA spinwave results discussed in the 
previous section. We have shifted the spinwave results by the renormalization 
factor $Z_c = 1.18$ obtained using several different numerical methods
\cite{numz,sandvik} 
(this value is also in close agreement with the $1/S^2$ spinwave 
value $Z_c=1.177$\cite{canali}). Spin wave theory clearly correctly predicts 
the number of the dominant peaks, which hence can be characterized as 
two-magnon peaks. There are, however, some discrepancies in the peak positions
and their relative weights. Most notably, for the largest lattice 
($6 \times 6$), the separation between the two peaks in the exact spectrum 
exceeds by a factor of more than $1.5$ that of the spinwave result. This may 
well be an indication that the correct two-magnon profile is broader than 
the standard profile obtained with spinwave theory.\cite{parkinson,canali}
Whether or not it is as broad as that obtained by Chubukov and Frenkel 
(see Fig.~\ref{figprofiles}) cannot be determined from the results for 
these small lattices, however. We will return to this important issue in 
the next section.

It can be noted that for the $4\times 4$ lattice there is a small peak
at $\omega /J \approx 4.5$ both in the spinwave and the Lancz\"os results. 
This then suggests that it is a two-magnon peak, in contrast to previous
claims that it arises from a four-magnon state.\cite{diag} For the larger 
lattices there is visible weight extending up to $\omega/J \approx 7$, which 
is not present in the spinwave results and hence is due to processes 
involving more than two magnons. For the $6\times 6$ lattice the relative 
weight of these contributions is about $10$\%.

It is difficult to scale the full shape of the spectrum to infinite system 
size. The first few frequency cumulants can be expected to converge rather 
quickly, however, and have also previously been calculated using a series 
expansion method as already discussed. The $n$th frequency {\it moment} is 
given by 
(at $T=0$)
\begin{equation}
\rho_n=\int_0^\infty d\omega \omega^n I(\omega).
\end{equation}
The first cumulant $M_1=\rho_1$, and for $n > 1$,
\begin{equation}
(M_n)^n={1\over \rho_0}
\int_0^\infty d\omega (\omega-\rho_1)^nI(\omega) .
\end{equation}
The results for $6 \times 6$ ($4\times 4$) are $M_1=3.524$ ($3.244$),
$M_2 = 0.8686$ ($0.797$), and $M_3=0.9576$ ($1.141$). The previous
series expansion results \cite{singh} are $M_1=3.58 \pm 0.06$, 
$M_2 = 0.81 \pm 0.05$,
and $M_3 = 1.00 \pm 0.14$. Hence, the $6 \times 6$ cumulants show an
improved and good agreement with the series results. However,
since there are significant differences between the $4\times 4$ and the 
$6 \times 6$ lattices, the results may still change in the thermodynamic 
limit. Unfortunately, using also the results for $N=26$ and $32$, the data 
do not fall on smooth curves (see Fig.~\ref{figcumulants} in the next 
Section), and it is not possible to extrapolate to the thermodynamic limit 
using only these Lancz\"os results. In the next Section we will calculate the 
cumulants for much larger systems using QMC data.

\section{Quantum Monte Carlo and Maximum-Entropy Analytic
Continuation}

Real-frequency dynamic properties cannot be obtained directly using QMC 
methods. Instead, the corresponding imaginary-time dependent correlation 
function has to be calculated, and numerically continued to real frequency. 
For the Raman spectrum defined by Eq.~(\ref{fermi}), the imaginary-time 
function is given by
\begin{equation}
G(\tau) = \langle \hat H_{LF} (\tau) \hat H_{LF} (0) \rangle ,
\label{gtau}
\end{equation}
where $\hat H_{LF} (\tau) = {\rm e}^{\tau \hat H} \hat H_{LF} 
{\rm e}^{-\tau \hat H}$.
The analytic continuation to real frequency amounts to inverting 
the integral relation
\begin{equation}
G(\tau) = {1\over \pi} \int_{-\infty}^\infty d\omega
I(\omega) {\rm e}^{-\tau \omega} .
\label{gtaw}
\end{equation}
With $G(\tau)$ obtained only to within a statistical error from a QMC 
simulation, the spectrum $I(\omega)$ cannot be uniquely determined. In the 
Max-Ent approach to this difficult problem,\cite{maxent1,maxent2} a unique 
solutions is defined as that minimizing
\begin{equation}
Q = {1\over 2}\chi^2 - \alpha S,
\label{mentq}
\end{equation}
where $S$ is the entropy of the spectrum,
\begin{equation}
S = - \int_{-\infty}^\infty d\omega I(\omega)\ln{[I(\omega)/m(\omega)]},
\end{equation}
defined with respect to a ``default'' model $m$ (both $I$ and $m$ are here
assumed to be normalized to unity). $G(\tau)$ is calculated for a discrete
set of times $\tau_i$. A given $I(\omega)$ corresponds to unique values 
of $G(\tau_i)$ according to Eq.~(\ref{gtaw}). The deviation from the
actual calculated $G_{\rm QMC}(\tau_i)$ is quantified by $\chi^2$. Since the 
statistical errors $\sigma_i$ of $G_{QMC}(\tau_i)$ at different times are 
correlated (see Fig.~\ref{figexact4} in Appendix A), $\chi^2$ should be 
defined in terms of the inverse of the full covariance matrix $C$,
\begin{equation}
\chi^2 = \sum\limits_{i,j} 
[G(\tau_i)-G_{QMC}(\tau_i)\bigr ]C^{-1}_{ij}
[G(\tau_j)-G_{QMC}(\tau_j)\bigr ].
\label{chi2}
\end{equation}
We here parametrize the spectrum in terms of $N_\omega \sim 200-400$ equally 
spaced delta-functions $\delta (\omega-\omega_i)$ for $\omega_i > 0$:
\begin{equation}
I(\omega) = \sum\limits_{i=1}^{N_\omega} I_i \delta (\omega_i-\omega),
\end{equation}
and a smooth continuous spectrum is then represented by the curve
connecting the amplitudes $I_i$ (or by giving the $\delta$-functions a width
of the order of the frequency spacing, which gives a very similar curve).
The negative part of the bosonic spectrum is given by detailed balance: 
$I(-\omega)= {\rm e}^{-\beta\omega}I(\omega)$. We use a flat default model 
for $\omega > 0$. The parameter $\alpha$ in Eq.~(\ref{mentq}) is determined 
iteratively so as to satisfy the ``classic'' Max-Ent criterion, resulting
in (within the assumptions of the Max-Ent method) the spectrum with the 
highest probability given the QMC data.

For calculating $G(\tau_i)$ we use the stochastic series expansion
QMC method,\cite{qmc1,qmc2} as discussed in Appendix A. With this method, 
derivatives of $G(\tau)$ can also be directly calculated. We here use the 
first two derivatives as supplementary information in the Max-Ent method. 
The $n$th derivative of $G(\tau)$ is related to $I(\omega)$ according to
\begin{equation}
G^{(n)}(\tau) = {(-1)^n\over \pi} \int_{-\infty}^\infty 
d\omega \omega^n I(\omega) {\rm e}^{-\tau\omega} .
\label{gntau}
\end{equation}
It is a straight-forward matter to modify the Max-Ent procedures to
include also the first few (in our case 2) of these derivative relations
in addition to the original analytic continuation equation (\ref{gtaw}).
The use of derivatives was first suggested by Sch\"uttler and Scalapino 
in their pioneering work on numerical analytic continuation based on 
$\chi^2$ fitting to QMC data.\cite{schuttler} To our knowledge, the Max-Ent 
method has not previously been used  with derivative information. It should 
be noted that the $n$th frequency moment $\rho_n$ is given by the $\tau=0$ 
derivative:
\begin{equation}
\rho_n = (-1)^n {G^{(n)} (\tau \to 0) \over G(\tau \to 0)} .
\label{gratio}
\end{equation}
Enforcing known frequency moments has been previously used to improve the 
resolution of the Max-Ent method.\cite{whitemoments}  The derivative
information goes beyond this by enforcing also the ``moments'' defined
with $\tau > 0$ in Eq.~(\ref{gratio}). The derivatives can of course be 
expected to improve on the Max-Ent procedure only if they can be calculated 
accurately enough to contain information not already present in the calculated
$G(\tau)$. Typically, the statistical errors increase with increasing
derivative order $n$. In our case, the first two derivatives appear to be 
useful, although spectra obtained with only $G(\tau)$ are not dramatically 
different.

Next, we present results for systems of size $L\times L$, with $L=4$, $6$, 
$8$, and $10$. Although considerably larger lattices can be studied with 
the QMC method, the physical information we are interested in here requires 
very accurate results for $G(\tau)$. It is therefore more appropriate to 
concentrate the computational resources on obtaining reliable results for 
moderate system sizes. Comparing results for $L=4-10$ should also be 
sufficient for making statements about the thermodynamic limit. We also
carried out some simulations for $L=16$, but the statistical error are
significantly larger in this case and the continuation to real frequency
is therefore less reliable. In order to obtain ground state results, the
simulations were carried out at inverse temperatures as high as $\beta=8L$. 
Results obtained with $\beta=4L$ are indistinguishable within statistical 
errors, indicating that contributions from excited states indeed are 
negligible at these low temperatures.

We begin by showing in Fig.~\ref{figltau} our results for the logarithm of 
the normalized imaginary-time correlator $g(\tau)=G(\tau)/G(0)$. In the same
figure we also show the relative statistical error, $\sigma_{\rm rel}(\tau)$,
of $g(\tau)$. Since the results for all the system sizes have comparable
errors, one can expect the Max-Ent continuation to real frequency to resolve 
structure on roughly  the same scale. Already from this 
imaginary-time data it is clear that the real-frequency spectum has dominant
weight at $\omega \approx 3J$ for all system sizes, as $\ln{[g(\tau)]}$
decays approximately linearly with $\tau$ in a sizable regime, with slope 
$\approx -3$. For the larger systems a slight upward curvature can be noted,
indicating that there is spectral weight also below $3J$. 

We find that the shape of the Raman spectrum obtained with the Max-Ent 
method is very sensitive to the statistical fluctuations in the QMC data.
Carrying out the Max-Ent procedures with different subsets of the available 
imaginary-time data always gives a dominant peak close to $\omega/J=3$, 
but the peak width and asymmetry show large variations. We therefore 
consider it appropriate to define the spectrum corresponding to the full 
set of imaginary time data as an average over suitably defined subsets. 
For this purpose we use the so called bootstrap method \cite{bootstrap} 
in the following way.

With the simulation data for some quantity $A$ divided into $M$ ``bin 
averages'' $A_i$ in the standard way, a bootstrap sample $A_B$ is defined as
\begin{equation}
A_B = {1\over M} \sum_{i=1}^M A_{R_i},
\end{equation}
whre $R_i$ is a randomly chosen bin (i.e., the number of bins chosen is
the same as the total number of bins, allowing, of course, for multiple
selections the bins). Since the Max-Ent procedure is highly non-linear,
the average over a large number of separately Max-Ent continued bootstrap 
samples of imaginary-time correlation functions can be different from 
the continued full average. We argue that the bootstrap average is more 
meaningful since statistical fluctuations are averaged out considerably. 

In Fig.~\ref{figbootstraps} we show Max-Ent results for 10 bootstrap samples
of $4\times 4$ QMC data. All the spectra have a dominant peak very close
to the correct position $\omega/J = 2.98$, as well as a structure at higher
frequency. There are, however, very large variations in the peak width
and in the position of the high-frequency weight. The average over 500
bootstrap samples is shown in Fig.~\ref{figmaxent}. The exact Lancz\"os
result with a damping $\epsilon/J=0.1$ is quite well reproduced, except
that the small peak at $\omega/J = 4.5$ is not present. It can be noted
that the main peak of the average spectrum is narrower than most of
the ``typical''  bootstrap samples (see Fig.~\ref{figbootstraps}), 
contrary to what might have been expected. This is clearly due to the
fact that the position of the peak shows very small variations compared
to the variations in the peak width and that for some bootstrap samples
the peak is very sharp.

Results for the larger lattices are also shown in Fig.~\ref{figmaxent}. In 
the $6\times 6$ spectrum the two main peaks are clearly resolved. 
The weight present 
at higher frequency cannot be resolved as a separate structure, however, 
and instead causes the shift by $\approx 15$\% of the second peak. As the 
system size grows, the number and density of peaks increase, and only a 
single structure can then be resolved. For the 
$8 \times 8$ lattice, the spinwave result shown in Fig.~\ref{figspinwave} 
has only one dominant peak. The Max-Ent result for this size is, however, 
very broad, indicating that the relative weight distribution among the peaks 
obtained in spinwave theory is not reliable (signs of this is seen also
in the exact $6\times 6$ spectrum in Fig.~\ref{figlanczos}). In particular, 
the Max-Ent spectrum has  much more low-frequency weight. This is the case 
also for the $10 \times 10$ lattice. The spectrum has a more pronounced peak 
than for $8 \times 8$, indicating that the individual $\delta$-functions 
begin to group into a profile peaked around $\omega /J \approx 3.5$. It 
should be noted that the procedures we are using can be expected to work 
better for the larger systems, for which the distribution of 
$\delta$-functions are better approximated by a single continuous 
structure. 

In Fig.~\ref{figderiv} we show the results for the short-time behavior of the
ratio $G^{(n)}(\tau)/G(\tau)$, along with the corresponding curves obtained 
from the Max-Ent results. According to Eq.~(\ref{gratio}), the first two
frequency moments can be directly obtained from the $\tau =0$ points. The 
first moment can be accurately extracted this way. In the case of the second 
moment the statistical fluctuations grow large as $\tau \to 0$, but the
extrapolation provided by the Max-Ent fit still gives a quite stable result. 
We also extract the third moment from the Max-Ent spectra. For both
$L=4$ and $L=6$ the results are in excellent agreement with the exact
results obtained with the Lancz\"os method in Sec.~III.
 
Fig.~\ref{figcumulants} shows the system size dependence of both the QMC 
and the Lancz\"os results for the cumulants, along with the previous 
\cite{singh}
infinite-size series results by Singh {\it et al.}. We also include the first
and second cumulants obtained for a $16 \times 16$ lattice, for which we do 
not consider the full line shape obtained with the Max-Ent method to be
stable due to larger statistical errors than for the smaller systems.
The first two cumulants can nevertheless be estimated. The Max-Ent and 
series results agree very well for the larger systems. The exact results for 
the non-square lattices do not show a regular size dependence, whereas the 
$L \times L$ lattices do. The {\it moments} for the $L \times L$ lattices
increase monotonically with $L$. However, there is a clear maximum in the 
second cumulant for $L=6$. This is likely caused by the lack of weight 
between the two dominant peaks for this lattice size. With growing size the 
gap should gradually be filled in by other peaks, leading to a decreasing 
second cumulant. Judging from Fig.~\ref{figcumulants}, the results for
the largest systems ($16 \times 16$ for $M_1$ and $M_2$ and $10 \times 10$
for $M_3$) should represent the thermodynamic limit within statistical 
errors. We then have $M_1 = 3.59 \pm 0.01$, $M_2 = 0.79 \pm 0.03$, and
$M_3 = 0.95 \pm 0.08$.

We now return to the line shape. The Max-Ent spectra displayed in 
Fig.~\ref{figmaxent}  show a considerable 
dependence on the lattice size. The trend for $L \ge 6$ appears to be
the development of a well defined main peak at $\omega/J \approx 3.5$, as 
well as some strengthening of the tail up to $\omega/J \approx 7$. Comparing
with the spinwave results for the two-magnon profiles shown in 
Fig.~\ref{figprofiles}, the $10 \times 10$ Max-Ent spectrum is clearly much 
broader than the narrow peak obtained by Canali and Girvin,\cite{canali}
but not quite as much broadened towards lower frequencies  as the 
Chubukov-Frenkel profile \cite{chubukov}
obtained by setting $S=1/2$ at a later stage of the calculation. Since the 
Max-Ent method can be expected to cause some broadening and the trend with 
increasing the lattice size appears to be a narrowing of the dominant peak, 
we conclude that the actual peak in the  thermodynamic limit should be narrower
than that obtained by Chubukov and Frenkel. 

In the exact $6 \times 6$ result 
there are contributions in the frequency range $\omega /J \approx 4.5-7$ 
which are not present in the spinwave result for the same lattice (see 
Fig.~\ref{figlanczos}). This weight is therefore most likely dominated by
processes involving more than two magnons. The Max-Ent result for the 
$10\times 10$ lattice also shows a tail extending up to $\omega /J \approx 7$.
The total weight above the spinwave theory two-magnon cut-off $\omega/J = 
4.63$ does, however, remain at $\approx 10-15$\%, as previously argued on 
the basis of the $1/S^2$ spinwave results and the frequency 
cumulants.\cite{canali}

Canali and Girvin argued that the two-magnon profile is very little
affected by the higher-order processes, and that the four-magnon
contribution should be a peak well separated from the two-magnon profile. 
\cite{canali} We now consider an approach to testing this hypothesis 
numerically, independently of the Max-Ent method. We assume a spectrum 
consisting of the Canali-Girvin two-magnon profile $P(\omega)$ shown in 
Fig.~\ref{figprofiles}, and a Gaussian $G_{\sigma_4}(\omega-\omega_4)$ of 
width $\sigma_4$ centered at $\omega=\omega_4$ for modeling the 
higher-order contribution. In order to account for a possible  further 
frequency shift, we use a phenomenological frequency renormalization 
$Z$ in the two-magnon profile. The full spectrum is hence
\begin{equation}
I(\omega) = A_2 P(Z\omega) + A_4 G_{\sigma_4}(\omega-\omega_4),
\end{equation}
where $P$ and $G_{\sigma_4}$ are both normalized to one, and hence 
$A_2+A_4=1$. We then have four parameters; $Z$, $A_4$, $\omega_4$, and 
$\sigma_4$ which can be adjusted to give the best consistency with the 
imaginary-time data. Note that $P(\omega)$ already contains the spinwave 
renormalization factor to order $1/S^2$, and hence our $Z$ should be close 
to $1$ for this treatment to be consistent. 

For the $10\times 10$ lattice, the imaginary-time data can indeed be very 
well accounted for by this spectrum. We obtain the parameters $Z=0.97$, 
$A_4=0.40$, $\omega_4=4.1$ and $\sigma_4=1.1$. The resulting spectrum
is shown in Fig.~\ref{figexperiment}. The data for $16 \times 16$ spins
can also be very well fit to the form considered, and the parameters are not 
changed much from the $10\times 10$ ones. This spectrum is also shown in 
Fig.~\ref{figexperiment}. The parameters of the Gaussian are such that it 
is completely merged with the two-magnon profile. This is clearly consistent 
with both the $6\times 6$ Lancz\"os and the  Max-Ent results, which did not 
show any significant gap between the main peak and the high-frequency 
weight. In Fig.~\ref{figexperiment} the weight of the Gaussian also extends 
to the low-frequency side of the two-magnon peak, and therefore has the 
effect of broadening it. Therefore, the relative weight of about $40$\% 
of the secondary peak cannot be interpreted directly as the total 
four-magnon (and higher)
contribution, but also likely reflects that the two-magnon profile 
from spinwave theory is too narrow. We have also carried out fits to two 
Gaussians, and then find that the dominant one is at a position 
$\approx 3.2J$, and the second one again is at $\approx 4-4.5J$. However, 
the uncertainty in the width of the dominant peak is large, and therefore 
this method cannot be used to accurately determine the width. Based on the 
other approaches we have discussed, we can nevertheless conclude that the 
standard spinwave two-magnon profile is too narrow, but by how much is not 
completely clear. The profile shown in Fig.~\ref{figexperiment} likely 
represents a lower bound of the width.

We also attempted a similar fitting procedure using the Chubukov-Frenkel
two-magnon result as the dominant feature. However, we found that it was
not possible to obtain any good fit to the QMC data in this case, due
to the, apparently, too high low-energy weight.

In Fig.~\ref{figexperiment} we also show an experimental spectrum for
La$_{\rm 2}$CuO$_{\rm 4}$, with the frequency scale adjusted to give the 
same peak position $\omega /J \approx 3.25$ as the QMC-spinwave fit. This peak 
position corresponds to an exchange constant $J = 1440$ K for the
experimental system, which is in good agreement with $J \approx 1500$ K
obtained from Neutron scattering and NMR experiments. Although the experimental
spectrum is somewhat broader than our result, there is a quite good agreement 
with the distribution of the weight present above the two-magnon cut-off 
frequency. Since the width of the theoretical spectrum shown here most 
probably is a lower bound of the actual width, we do not consider the 
deviations from the experimental spectrum serious. Comparing with the 
two-magnon spinwave spectra shown in Fig.~\ref{figprofiles}, it is clear 
that our present fitted spectrum is considerably closer to the experimental 
result. As will be discussed further in Sec.~VI, the  width of the peak is 
such that the further broadening required to match the 
experimental spectrum could quite easily be achieved by spin-phonon 
couplings, as has been suggested by several groups.

\section{Finite-temperature results}

In this section we present results of QMC and Max-Ent calculations 
carried out at temperatures $T/J =0.25$, $0.5$, and $1.0$.\cite{previous}
Raman spectra for a $4\times 4$ lattice at these temperatures were previously 
obtained by Bacci and Gagliano using exact diagonalization.\cite{bacci} 
Recently, finite temperature Lancz\"os calculations for lattices with up to 
$20$ sites were presented by Prelov\v{s}ek and Jakli\v{c}.\cite{prelovsek} 
Here we compare QMC+Max-Ent results for systems with $4\times 4$ and 
$16 \times 16$ spins. The latter size should be sufficient for obtaining 
thermodynamic limit results at the temperatures considered.

Fig.~\ref{figtempgt} shows the imaginary-time correlation functions. For the
temperatures considered here, $g(\tau)$ can be accurately evaluated
for the whole range $0 \le \tau \le \beta$. The slower decay with $\tau$
for the larger lattice indicates the presence of more low-frequency
weight as the system size increases. This is confirmed by the Max-Ent 
results for the real-frequency spectra, 
shown in Fig.~\ref{figtspecs}. The results for 
$4\times 4$ spins are in reasonable agreement with exact diagonalization
 results if one includes some rather large broadening of the 
$\delta$-functions. In Fig.~\ref{figtspecs} we have graphed the exact 
results as histograms, with the bin width for each temperature chosen large 
enough to remove most, but not all, of the jagged structure due to the 
discrete finite-size spectrum. It is clear that the Max-Ent method cannot 
capture the fine-structure of the spectrum, and instead gives a single
rounded shape. Nevertheless, the region of dominant spectral weight
and its temperature variations are well reproduced. The low-frequency peak 
in the exact $4\times 4$ spectra at high temperatures is due to degeneracies 
present for this small lattice \cite{bacci} (i.e., the peak is actually 
at $\omega = 0$).  

Our $16 \times 16$ results show a faster enhancement of the low-frequency 
spectral weight as the temperature is increased above $T/J \approx 0.25$. 
This difference between lattice sizes is likely due to the presence of 
large finite-size gaps in the level spectrum of the $4 \times 4$ system. 
Naturally, as $T \to \infty$ the system size dependence should diminish, 
and this is seen already at $T=1.0$ in Fig.~\ref{figtspecs}. The 
finite-temperature spectra calculated for $20$ sites by Prelov\v{s}ek and 
Jakli\v{c} \cite{prelovsek} for $T/J=1.0$ and $0.5$ are in reasonable 
agreement with our $16\times 16$ results, again taking into account a 
Max-Ent broadening of our spectra. However, at $T/J=0.5$, judging from 
the rather large differences between the exact results for $N=16$ 
(Ref.~\onlinecite{bacci}) and $N=20$ (Ref.~\onlinecite{prelovsek}) and the 
slow approach to the thermodynamic limit discussed in sec.~IV, it is 
likely that the $N=20$ spectrum has not yet converged to its infinite-size 
shape. The actual width at this temperature should therefore be something 
intermediate between our $16 \times 16$ Max-Ent result and the previously
obtained $N=20$ profile.

Experimentally, spectra taken at room temperature do not differ 
significantly from ones obtained at very low temperatures.
\cite{ramanexp1,knoll}  As the temperature is elevated to to 
$T/J \approx 0.5$ there is a significant 
increase in the weight below $\omega \approx 2J$.\cite{knoll} This feature 
is indeed quite well reproduced by our result for $16 \times 16$ spins. 

The spectra shown in Fig.~\ref{figtspecs} are all normalized to $1$. The
temperature dependence of the integrated intensity is a quantity of 
experimental interest. We define two intensities:
\begin{mathletters}
\begin{eqnarray}
I_1 & = & \int_{-\infty}^\infty d\omega A(\omega), \\
I_2 & = & \int_{0}^\infty d\omega A(\omega).
\end{eqnarray}
\end{mathletters}
These definitions are equivalent at $T=0$, but differ at finite 
$T$ due to spectral weight at negative frequencies, with
$A(-\omega)={\rm e}^{-\beta\omega}A(\omega)$. $I_1$ can
be directly obtained from the imaginary-time data as $G(\tau=0)$,
whereas $I_2$ is calculated by integrating the real frequency spectrum
obtained using the ME method. Fig.~\ref{figintens} shows both intensities
vs.~$T$ for $4\times 4$ and $16\times 16$ lattices. Up to $T/J \approx
0.25$, $I_1 \approx I_2$, owing to the absence of significant
low-frequency weight at these temperatures. At higher temperatures
$I_1 > I_2$, but even at $T/J \approx 0.5$ the difference is small.
For the $4\times 4$ system the intensity $I_2$
increases by $\approx 14 \%$ as the temperature is decreased from 
$T/J=0.5$ to $T/J=0$, and for $16 \times 16$ by $\approx 9\%$.

\section{Summary and discussion}

We have presented numerical results for the $B_{1g}$ spectrum of the 
Heisenberg model within Loudon-Fleury theory. We obtained Lancz\"os exact 
diagonalization results for up to $6 \times 6$ spins, and carried out
QMC simulations for up to $16 \times 16$ spins. We compared the results with
spin wave theory. Our main results and conclusions are the following: 

1) Comparing spinwave theory and exact diagonalization results for the same 
lattice sizes, we find that for a given cluster the number of dominant peaks 
is the same in both cases. 
However, both the positions of the peaks and their relative weights are 
different. Most notably, for the $6 \times 6$ lattice there are two dominant 
peaks, the separation of which is $1.5$ times larger in the exact result. 
Assuming that the trend persists for larger lattices, this indicates that 
spinwave theory underestimates the width of the dominant $B_{1g}$ peak in 
the thermodynamic limit. 

2) Our results of Maximum-Entropy analytic 
continuation of QMC imaginary-time data is also consistent with a peak 
width larger than that of the spinwave two-magnon peak. The first three 
frequency cumulants are in excellent agreement with previous results of a 
series expansion around the Ising model. We estimate the cumulants in the 
thermodynamic limit to be $M_1 = 3.59 \pm 0.01$, $M_2 = 0.79 \pm 0.03$, 
and $M_3 = 0.95 \pm 0.08$. 

3) In order to test the $1/S^2$ spinwave theory prediction of a four-magnon 
profile well separated from the main two-magnon peak,\cite{canali}
 we carried out a fit of 
the QMC imaginary-time data to a spectrum consisting of the spinwave 
two-magnon peak and a Gaussian at higher frequency. We found that this type 
of spectrum indeed describes the data well. The fitted Gaussian is centered at 
$\omega /J \approx 4.1$, and is so broad that it is completely merged 
together with the two-magnon structure peaked at $\omega /J \approx 3.25$. 
The resulting spectrum  resembles a typical experimental $B_{1g}$ 
profile for La$_{\rm 2}$CuO$_{\rm 3}$ with an exchange $J \approx 1400$ K. The
experimental peak is still slightly broader, but there is a considerable
improvement in comparison with the standard spinwave theory two-magnon 
profile. 

4) The imaginary-time data cannot be fitted using the two-magnon profile 
obtained by Chubukov and Frenkel \cite{chubukov}
by expanding their spinwave spectrum around
its peak position before setting $S=1/2$ in the calculation. This is
due to the significantly stronger low-frequency weight present in this
spectrum.

5) At finite temperature we find a significant increase in spectral
weight below $\omega \approx 2J$ for $T/J \agt 0.25$, in agreement with
experimental results for antiferromagnetic cuprates.\cite{knoll} We also 
find that this effect is suppressed in the $4 \times 4$ system, due to the 
finite-size gaps. The temperature dependence of the integrated scattering 
intensity is weak.

Our results hence confirm that LF theory can account for some of the 
main features of typical $B_{1g}$ spectra observed experimentally for 
antiferromagnetic cuprates such as La$_2$CuO$_4$. Our new evidence for a 
profile significantly broader than that obtained in spinwave theory support
in part the early claim by Singh {\it et al.} \cite{singh} that the 
broadening is due
to the strong quantum fluctuations of the Heisenberg model with 
$S=1/2$ (note that spinwave theory is in good agreement with experimental 
results for quasi-2D $S=1$ systems \cite{spin1}). However, 
typical experimental spectra are still  broader, and extend to slightly 
higher frequencies. The shoulder-like feature observed in some experiments 
at $\omega \approx 4J$ is also not present in our results, although we find 
evidence that the four-magnon contribution has its maximum in this regime. 
Hence, although our results show a better agreement with experiments than 
previous numerical results obtained for smaller lattices,\cite{diag,bacci} 
the Heisenberg-LF mechanism does not appear to fully account for the
experimental Raman scattering, as has been noted in several previous 
studies. The fact that there is no $A_{1g}$ scattering within this theory 
of course also implies that other additional mechanisms have to be active. 

Chubukov and Frenkel recently suggested that resonant processes not
contained within LF theory are important in typical experiments, for
which the frequency $\omega_{\rm in}$ of the incident light is comparable 
to the charge transfer gap of the CuO$_{\rm 2}$ planes.\cite{chubukov} We 
agree that resonance effects are most likely needed to explain the
dependence of the total scattering intensity and the line 
shape on $\omega_{\rm in}$,\cite{blumberg} but note that the 
dominant features of the profile do not show much dependence on 
$\omega_{\rm in}$ for most of the frequencies studied.
\cite{ramanexp1,blumberg} 
Based on the improved agreement with experiments obtained here within LF 
theory, we believe that the main features of the $B_{1g}$ spectrum are due 
to the LF mechanism, and that a further broadening of the spectrum could 
be achieved by magnon damping due to phonons.

Motivated by experiments carried out at high temperatures, Knoll 
{\it et al.}~suggested that spin-lattice interactions may be responsible
for the broadening of the Raman spectrum.\cite{knoll} Spinwave 
calculations including a phenomenological magnon life-time give some
support to these ideas.\cite{weber} Several different calculations
explicitly including magnon-phonon coupling have been presented 
recently.\cite{nori,manning,saenger} Using an adiabatic approximation for 
the phonons leads to a Heisenberg model with random coupling constants. 
Numerically studying such random lattices with $4\times 4$ spins and 
assuming a standard LF coupling, Nori {\it et al.} found that the 
$B_{1g}$ spectrum can indeed be broadened by this mechanism, and that 
also $A_{1g}$ scattering can become significant.\cite{nori} However,
in this calculation, the strength of the randomness required in order 
to reproduce the width of the experimental $B_{1g}$ spectrum appears to 
be rather large (using a Gaussian distribution for the nearest-neighbor 
couplings $J_{ij}$, a width $\sigma \approx 0.5 \langle J_{ij} \rangle$ was 
required).\cite{norinote} Nori {\it et al.} argued that such strong disorder 
can be caused by incoherent atomic displacements. Nevertheless, in the absence 
of other evidence for the presence of large fluctuations in the Heisenberg
couplings, it would be desirable to reproduce the broadened spectrum with a 
narrower coupling distribution. 

One reason for the strong disorder required in the calculation of Nori 
{\it et al.} could be the small size of the lattice used.\cite{nori} As
we have seen, the pure $4\times 4$ system only has a single dominant 
two-magnon $\delta$-function at $\omega = 2.98J$, and two weaker peaks
at $\omega \approx 4.5J$ and $\omega \approx 5.5J$.\cite{diag}
It is clear that a considerably weaker disorder would suffice to broaden the 
spectrum if one starts from the much broader pure-system profile
obtained here for larger lattices. 

The type of QMC and Max-Ent calculations presented here 
could in principle be carried out also for disordered spin systems,
and even including fully dynamic 
phonons.\cite{phononqmc} The suggested effects 
of magnon-phonon coupling could hence be investigated more rigorously than 
previously, using larger lattices. Although a recent exact diagonalization 
study by Reilly and Rojo\cite{rr} give some support for the validity of an 
adiabatic approximation for the phonons, calculations with full dynamic 
phonons should also be carried out for larger lattices. Limits on the 
strength of the phonon-magnon coupling (or the width of the disorder 
distribution in the adiabatic approach) could be established by carrying 
out QMC calculations of, e.g., the temperature dependence of the spin 
correlation length \cite{ding} and NMR relaxation rates \cite{qmcrates} 
for systems including lattice vibrations or static disorder. 

\vskip2mm

\acknowledgments

We would like to thank G. Blumberg, C. Canali, A. Chubukov, S. Girvin, 
D. Morr, R. Singh, and P. Prelov\v{s}ek for useful discussions and 
correspondence. This work was supported by the National Science Foundation 
under Grants No.~DMR-95-20776, DMR-95-27304 and -DMR-97-12765. The QMC 
calculations were carried out at the Supercomputer Computations Research 
Institute at Florida State University.

\appendix

\section{QMC calculations of the Raman correlation functions}

In this Appendix we describe the calculation of the imaginary-time correlation
function (\ref{gtau}) with the stochastic series expansion (SSE) method.
\cite{qmc1,qmc2} In order to reduce the statistical fluctuations, we use the
spin-rotational invariance of the Heisenberg Hamiltonian to construct an 
estimator less noisy than the obvious one. We also derive direct estimators 
for the $\tau$-derivatives of $G(\tau)$. In order to establish the notation,
we first very briefly outline the formalism of the SSE algorithm. More 
details of the implementation of this non-standard generalization of 
Handscomb's method \cite{handscomb,lee} for the 2D Heisenberg model 
can be found in Ref.~\onlinecite{sandvik}.

In order to apply the SSE technique, the Hamiltonian is first written as
\begin{equation}
\hat H = -{J\over 2} \sum\limits_{b=1}^{2N} [\hat H_{1,b} - \hat H_{2,b}]
+ {NJ\over 2},
\label{ham2}
\end{equation}
where $b$ is a link connecting a pair of nearest-neighbor sites
$\langle i(b),j(b)\rangle$, and the operators $\hat H_{1,b}$
and $\hat H_{2,b}$ are defined as
\begin{mathletters}
\begin{eqnarray}
\hat H_{1,b} & = & 2[\hbox{$1\over 4$} - S^z_{i(b)}S^z_{j(b)}] \\
\hat H_{2,b} & = & S^+_{i(b)}S^-_{j(b)} + S^-_{i(b)}S^+_{j(b)} .
\end{eqnarray}
\end{mathletters}
An exact expression for an operator expectation value 
\begin{equation}
\langle \hat A \rangle = {1\over Z}
{\rm Tr}\lbrace \hat A {\rm e}^{-\beta \hat H} \rbrace ,\quad
Z = {\rm Tr}\lbrace {\rm e}^{-\beta \hat H} \rbrace ,
\end{equation}
at inverse temperature $\beta = J/T$, is obtained by Taylor expanding
exp$(-\beta \hat H)$ and writing the traces as  sums over diagonal matrix
elements in the basis $\lbrace |\alpha \rangle \rbrace =
\lbrace |S^z_1,\ldots,S^z_N \rangle \rbrace$. The partition function then
takes the form \cite{qmc1}
\begin{equation}
Z = \sum\limits_\alpha \sum\limits_n \sum\limits_{S_n}
{(-1)^{n_2}\over n!} \Bigl ( {\beta \over 2} \Bigr )^n 
\Bigl \langle \alpha \Bigl | 
\prod\limits_{l=1}^n \hat H_{a_l,b_l} \Bigr | \alpha \Bigr \rangle ,
\label{partition}
\end{equation}
where $S_n$ is a sequence of index pairs
defining the operator string $\prod_{l=1}^n \hat H_{a_l,b_l}$,
\begin{equation}
S_n = [a_1,b_1][a_2,b_2]\ldots [a_n,b_n],
\label{sn}
\end{equation}
with $a_i \in \lbrace 1,2\rbrace$, $b_i \in \lbrace 1,\ldots ,2N \rbrace$,
and $n_2$ denotes the total number of index pairs (operators) 
$[a_i,b_i]$ with $a_i = 2$. Both $\hat H_{1,b}$ and $\hat H_{2,b}$ can 
act only on states where the spins at sites $i(b)$ and $j(b)$ are 
antiparallel. $\hat H_{1,b}$ leaves such a state 
unchanged, whereas $\hat H_{2,b}$ flips the spin pair. Defining a 
propagated state
\begin{equation}
| \alpha (p) \rangle = \prod\limits_{l=1}^p \hat H_{a_l,b_l} |\alpha \rangle ,
\quad | \alpha (0) \rangle = | \alpha \rangle ,
\label{propagated}
\end{equation}
a contributing $(\alpha ,S_n)$ must clearly satisfy the periodicity
condition  $|\alpha (n) \rangle = | \alpha (0) \rangle$. In an allowed 
sequence $S_n$, the links $b$ corresponding to the spin-flipping operators 
$[2,b]$ present must therefore form only closed loops. For a lattice with 
$L \times L$ sites and $L$ even, this implies that the number $n_2$ 
must be even, and hence that all terms in Eq.~(\ref{partition}) are 
positive and can be used as relative probabilities in a Monte Carlo
algorithm (this is true for any non-frustrated system).
Since any non-zero matrix element in (\ref{partition}) is equal to 
one, the weight factor corresponding to a contributing $(\alpha ,S_n)$ 
is simply given by
\begin{equation}
W(\alpha,S_n) = {(\beta/2)^2 \over n!}.
\label{wn}
\end{equation}
The algorithm for sampling the configurations $(\alpha ,S_n)$ is described
in Ref.~\onlinecite{sandvik}. 

In order to obtain an expression for $G(\tau)$ in terms of the
states $|\alpha (p)\rangle$ and the index sequence $S_n$ used in the
simulation, the expectation value is first written in terms of the 
operators $\hat H_{a,b}$ as
\begin{equation}
G(\tau) = \sum\limits _{a_1,a_2}\sum\limits _{b_1,b_2} P_{b_1,b_2} 
G^{a_1,b_1}_{a_2,b_2} (\tau),
\end{equation}
where
\begin{equation}
G^{a_1,b_1}_{a_2,b_2} (\tau) = 
\langle H_{a_2,b_2} (\tau) H_{a_1,b_1} (0) \rangle,
\label{gsum}
\end{equation}
and $B_{1g}$ symmetry corresponds to $P_{b_1,b_2}=1$ for links $b_1$ and 
$b_2$ which are parallel to each other, and $P_{b_1,b_2}=-1$ for perpendicular
links. Proceeding as in the derivation of the partition function 
(\ref{partition}), the exponentials in the expression
\begin{eqnarray}
G&&^{a_1,b_1}_{a_2,b_2} (\tau) = \nonumber \\
&&{1\over Z} \sum\limits_\alpha
\bigl \langle \alpha \bigl | {\rm e}^{-(\beta -\tau)\hat H} 
\hat H_{a_2,b_2} {\rm e}^{ -\tau\hat H} \hat H_{a_1,b_1} 
\bigr | \alpha \bigr \rangle
\label{gabtau}
\end{eqnarray}
are Taylor expanded and all powers of $\hat H$ are written as sums
of products of the operators $\hat H_{a,b}$. There is then a one-to-one
correspondence between the terms in $G^{a_1,b_1}_{a_2,b_2} (\tau)$ and
Eq.~(\ref{partition}). Dividing out the factor
corresponding to the configuration weight (\ref{wn}) gives the average 
in the form of a function of $S_n$:
\begin{equation}
G^{a_1,b_1}_{a_2,b2}(\tau) =
\left\langle \sum\limits_{m=0}^{n-2} F(\tau,n;m)
N^{a_1,b_1}_{a_2,b_2}(m) \right\rangle ,
\label{tauoff}
\end{equation}
where
\begin{equation}
F(\tau,n;m) = {\tau^m (\beta -\tau )^{n-m-2}(n-1)!
\over \beta ^{n} (n-m-2)!m!},
\label{ftau}
\end{equation}
and $N^{a_1,b_1}_{a_2,b_2}(m)$ is the number of times the operators
$[a_1,b_1]$ and $[a_2,b_2]$ occur in $S_n$ (in the given order)
separated by $m$ other
operators. Hence, measuring $G^{a_1,b_1}_{a_2,b2}(\tau)$ simply
amounts to finding all pairs of operators $[a_1,b_1]$ and $[a_2,b_2]$
in the sequence $S_n$. The contribution to Eq.~(\ref{gabtau}) of 
each pair is a function of the relative separation of the operators, 
given by Eq.~(\ref{ftau}).

In order to obtain a simple expression 
for the full correlation function $G(\tau)$ it is useful to introduce 
a function $X(p)$, such that $X(p)=+1$ if the $p$:th operator 
in $S_n$ acts on a link in the $x$-direction, and $X(p)=-1$ if
it acts on a $y$ link. Numbering the bonds such that
$0 \le b \le N$ correspond $x$-bonds, and $N+1 \le b \le 2N$ 
correspond to $y$-bonds, the definition is hence
\begin{equation}
X(p) = \left \lbrace 
\begin{array}{cc}
+1, & b_p \le N \\
-1,  & ~b_p > N .
\end{array} \right.
\end{equation}
Eqs.~(\ref{gsum}) and (\ref{tauoff}) then give
\begin{equation}
G(\tau) = 
\biggl\langle \sum\limits_{p=1}^n \sum\limits_{m=1}^{n-1}
F(\tau,n,m-1) X(p)X(p+m) \biggr\rangle ,
\label{gtau1}
\end{equation}
where of course $X(p)$ is periodic; $X(n+1)=X(1)$.

In practice the estimator (\ref{gtau1}) is rather noisy. An 
improved estimator can be constructed as follows. First, the function
$X(p)$ is written as a sum of two terms, 
\begin{equation}
X(p)=X_1(p)+X_2(p),
\end{equation}
where $X_t(p)=\pm 1$ ($t=1,2$) for $x$ and $y$ bonds, as before, 
if the $p$:th operator in $S_n$, $[a_p,b_p]$, has $a_p=t$, but
$X_t(p)= 0$ if $a_p \not= t$. Hence
\begin{equation}
X_t(p) = \left \lbrace 
\begin{array}{ccc}
+1 & ,a_p=t, & b_p \le N \\
-1 & ,a_p=t, & b_p > N  \\
0  & ,a_p\not=t & 
\end{array} \right.
\end{equation}
If $a_p=1$, $X_1(p)$ can be averaged over all $2N$ choices of operators 
$[1,b]$ at position $p$. The weight $W(\alpha,S'_n)$ corresponding to a 
sequence $S'_n$ obtained by replacing the current operator $[1,b_p]$ at $p$ 
in $S_n$ is equal to the current weight $W(\alpha,S_n)$ if the 
corresponding spins at sites $i(b)$ and $j(b)$ are antiparallel in the 
propagated state $|\alpha (p)\rangle$, and is zero otherwise. Hence, 
$X_1(p)$ can be redefined as 
\begin{equation}
X_1(p) = \left \lbrace 
\begin{array}{cc}
[ N^A_x(p)-N^A_y(p)]/2N & ,a_p =1 \\
0 & ,a_p\not=1 
\end{array} \right.
\end{equation}
where $N^A_\gamma (p)$ is the number of antiparallel nearest-neighbor 
spin pairs in the $\gamma$-direction in 
$| \alpha (p) \rangle$. One can easily verify that this averaged
estimator can be used in products with both $X_1$ and $X_2$.
Hence, improved estimators for $X(p)X(p+m)$ in Eq.~(\ref{gtau1})
can be used for the terms $X_1X_1$, $X_1X_2$, and $X_2X_1$. For $X_2(p)$
no simple re-definition in terms of single-operator averaging 
can be constructed (replacing a single operator $[2,b]$ with any
other operator always leads to a non-contributing term), and hence 
the $X_2X_2$ contribution to (\ref{gtau1}) remains noisy. However, 
the rotational invariance of the Heisenberg 
Hamiltonian implies that,
\begin{eqnarray}
\bigl\langle X_2(p)&&X_2(p+m)\bigr\rangle
=\bigl\langle 2X_1(p)X_1(p+m) \bigr\rangle + \nonumber \\
&&\hbox{$1\over 2$} \bigl \langle X_1(p)X_2(p+m)+X_2(p)X_1(p+m)
\bigr \rangle ,
\end{eqnarray}
and therefore the $X_2X_2$ term does not even have to be evaluated. The 
final result for the improved estimator for $G(\tau)$ is hence
\begin{eqnarray}
G(\tau) = & &
\biggl\langle \frac{3}{2}\sum\limits_{p=1}^n \sum\limits_{m=1}^{n-1}
F(\tau,n;m-1) \Bigl [ 2X_1(p)X_1(p+m)+ \nonumber \\
& & X_1(p)X_2(p+m)+X_2(p)X_1(p+m) \Bigr ] \biggr\rangle .
\label{gtau2}
\end{eqnarray}
It should be noted that the function $F(\tau,n;m)$ is sharply 
peaked around $m \approx n\tau/\beta$ for large $\beta$, so that 
typically only a small fraction of the terms in (\ref{gtau2}) actually 
have to be evaluated. 

Eq.~(\ref{gtau2}) is valid for any $0 \le \tau \le \beta$, and the 
$\tau$-dependence appears only in the function $F(\tau,n;m)$. 
In contrast to standard 
Trotter-based QMC methods, the method discussed here can therefore 
be used to directly calculate also $\tau$-derivatives of imaginary-time 
dependent correlation functions. An expression for the $n$:th derivative
of $G(\tau)$ is simply obtained by replacing $F$ in (\ref{gtau2}) 
by its $n$:th derivative:
\begin{eqnarray}
G^{(n)}&&(\tau) = {d ^n G(\tau) \over d\tau^n} =  \biggl\langle 
\frac{3}{2}\sum\limits_{p=1}^n \sum\limits_{m=1}^{n-1} \times \nonumber \\
& & \Bigl ( {d ^n F(\tau,n,m-1) \over d\tau^n} \Bigr )
\Bigl [ 2X_1(p)X_1(p+m)+  \nonumber \\
& & X_1(p)X_2(p+m)+X_2(p)X_1(p+m) \Bigr ] \biggr\rangle .
\label{gderiv}
\end{eqnarray}
As discussed in Sec.~IV,
derivatives can be used as supplementary information in a numerical analytic 
continuation to real frequency. The derivatives at $\tau=0$ 
are of special interest, as they are related to moments of the spectral 
function [see Eq.~(\ref{gratio})].

We end this Appendix with a demonstration that the simulation results for 
$G(\tau)$ are indeed free from systematic errors. Since the absolute Raman 
scattering intensity is not contained in the LF theory, the amplitude of 
$I(\omega)$, and hence of $G(\tau)$, is irrelevant, and instead of $G(\tau)$ 
one can consider the ratio
\begin{equation}
g(\tau)=G(\tau)/G(0).
\end{equation}
Fig.~\ref{figexact4} shows the QMC result for this quantity calculated on a
$4\times 4$ lattice, along with the exact result obtained from $I(\omega)$ 
calculated using exact diagonalization. The statistical error of the QMC 
result is in the fifth decimal digit, and there is excellent agreement with 
the exact result within this accuracy. The absence of detectable systematical
errors in the QMC result for $g(\tau)$ is hence confirmed. Since $g(\tau)$ 
decays exponentially, the relative statistical error grows rapidly with 
$\tau$, and for $\tau \agt 3$ accurate results can not be easily obtained. 
This is the case also for larger systems.

\vfill\eject

\begin{figure}
\caption{Spinwave theory results for the $B_{1g}$ Raman profile calculated
on small lattices with $L \times L$ sites. The dashed curves are the results 
with the interactions neglected, and the solid ones are with interactions
in the final states included at the RPA level. A damping $\epsilon=0.05J$
has been used to broaden the $\delta$-functions.}
\label{figspinwave}
\end{figure}

\begin{figure}
\caption{Spinwave theory results for the $B_{1g}$ two-magnon
profile in the thermodynamic limit, compared with the experimental 
spectrum for La$_{\rm 2}$CuO$_{\rm 4}$ discussed in 
Ref.~\protect{\onlinecite{singh}} (bold solid curve). The solid curve 
corresponds to Eq.~(\protect{\ref{specs1}}) 
with a spinwave renormalization factor $Z_c=1.158$. The dotted curve is the 
result by Canali and Girvin, which includes also quantum fluctuations in the 
ground state. The dashed line is the result by Chubukov and Frenkel, obtained
by further expanding the line shape (\protect{\ref{specs1}}) in $1/S$ before 
setting $S=1/2$. All curves are normalized to one. The frequency scale of the
experimental spectrum has been adjusted to give a peak position in rough 
agreement with the theoretical curves.}
\label{figprofiles}
\end{figure}

\begin{figure}
\caption{Exact diagonalization results for the $B_{1g}$ spectrum for different
small lattices with $N$ sites (solid curves). The dashed curves are the 
corresponding RPA-spinwave results. The $\delta$-functions of the exact 
results have been broadened using a damping $\epsilon=0.1J$, and all  the 
spectra ar normalized to one. The spinwave results have been given
a smaller damping and a different normalization in order to more clearly 
show the peak positions.}
\label{figlanczos}
\end{figure}

\begin{figure}
\caption{QMC results for $\ln{[g(\tau)]}$ (upper panel) for different system 
sizes, and the relative statistical errors of $g(\tau)$ (lower panel).}
\label{figltau}
\end{figure}

\begin{figure}
\caption{Results of Max-Ent analytic continuation of 10 bootstrap samples
of QMC imaginary-time data generated for a $4\times 4$ system.}
\label{figbootstraps}
\end{figure}

\begin{figure}
\caption{Bootstrap-averaged Max-Ent results for the $B_{1g}$ spectrum for
different lattices (solid curves). The $4\times 4$ and $6 \times 6$ results 
are compared with the corresponding exact diagonalization results with
a damping $\epsilon/J = 0.1$ (dotted curves).}
\label{figmaxent}
\end{figure}

\begin{figure}
\caption{QMC results for the short-time behavior of $G^{(1)}/G$ (upper 
panel) and $G^{(2)}/G$ (lower panel) for systems of linear sizes $L=4$ (solid
circle), $6$ (open circles), $8$ (solid squares), and $10$ (open squares). 
The solid curves are obtained from the Max-Ent analytic continuation.}
\label{figderiv}
\end{figure}

\begin{figure}
\caption{The first three frequency cumulants of the Max-Ent spectra vs.~the 
inverse system size (solid circles with error bars). The open circles 
are the exact diagonalization results. The previous infinite-size results from 
a series expansion, calculated by Singh {\it et al.},
\protect{\cite{singh}} are indicated by the 
horizontal dashed lines (result $\pm$ estimated error).}
\label{figcumulants}
\end{figure}

\begin{figure}
\caption{$B_{1g}$ spectrum obtained by a fit of imaginary-time QMC data to 
the Canali-Girvin two-magnon profile \protect{\cite{canali}} plus a 
Gaussian. The almost 
indistinguishable solid and dashed curves are for a $10 \times 10$ and a
$16 \times 16$ lattice, respectively. The bold curve is the experimental
spectrum for La$_{\rm 2}$CuO$_{\rm 4}$ discussed in 
Ref.~\protect{\onlinecite{singh}}, with the frequency scale adjusted to give 
the same peak position as the theoretical results (corresponding to
an exchange $J=1440 $ K).}
\label{figexperiment}
\end{figure}

\begin{figure}
\caption{The logarithm of the normalized imaginary-time correlator $g(\tau)$ 
vs.~$\tau$ for $4\times 4$ (dashed curves) and $16 \times 16$ (solid curves) 
lattices at different temperatures.}
\label{figtempgt}
\end{figure}

\begin{figure}
\caption{Max-Ent results for the $B_{1g}$ spectrum of $4\times 4$ (dashed 
curves) and $16 \times 16$ (solid curves) lattices at different temperatures.
The histograms represent the exact results for the $4\times 4$ lattice.}
\label{figtspecs}
\end{figure}

\begin{figure}
\caption{Integrated $B_{1g}$ scattering intensities vs. temperature for 
$4\times 4$ (open symbols) and $10 \times 10$ (solid symbols) lattices. 
Circles are for $I_1$ (using all frequencies), and squares for $I_2$
(using positive frequencies only).}
\label{figintens}
\end{figure}

\begin{figure}
\caption{Upper panel: QMC results for $g(\tau)=G(\tau)/G(0)$ of a $4\times 4$
system at inverse temperature $\beta=32$ (solid circles), compared
with the exact ground state result (solid curve). The inset shows
the regime $1.75 \le \tau \le 3$ on a more detailed scale. Lower panel: The
deviation of the QMC data from the exact result, multiplied by
$10^4$. The dashed curves indicate the statistical errors.}
\label{figexact4}
\end{figure}


\begin{references}

\bibitem{manousakis}
For a review, see E. Manousakis, Rev. Mod. Phys. {\bf 63}, 1 (1991).

\bibitem{neutrons1}
D. Vaknin {\it et al.}, Phys. Rev. Lett. {\bf 58}, 2802 (1987);
G. Shirane {\it et al.}, Phys. Rev. Lett. {\bf 59}, 1613 (1987);
Y. Endoh {\it et al.}, Phys. Rev. B {\bf 37}, 7443 (1988).

\bibitem{neutrons2} 
G. Aeppli {\it et al.}, Phys. Rev. Lett. {\bf 62}, 2052 (1989); 
S. M. Hayden {\it et al.}, Phys. Rev. B {\bf 42}, 10220 (1990);
S. M. Hayden {\it et al.}, Phys. Rev. Lett. {\bf 67}, 3622 (1991).

\bibitem{neutrons3} 
P. Bourges, H. Casalta, A. S. Ivanov, and D. Petitgrand,
preprint, cond-mat/9708060 (1997).

\bibitem{imai}
T. Imai {\it et al.} Phys. Rev. Lett. {\bf 70}, 1002 (1993); {\it ibid.},
{\bf 71}, 1254 (1993).

\bibitem{matsu}
M. Matsumura {\it et al.},
J. Phys. Soc. Jpn {\bf 63}, 4331 (1994).

\bibitem{csh} 
S. Chakravarty, B. I. Halperin, and D. R. Nelson, Phys. Rev. Lett.
{\bf 60}, 1057 (1988); Phys. Rev. B {\bf 39}, 2344 (1989).

\bibitem{ding} 
H.-Q. Ding and M. S. Makivi\'c, Phys. Rev. Lett. {\bf 64}, 1449 (1990);
M.~S.~Makivi\'c and H.-Q. Ding, Phys. Rev. B {\bf 43}, 3562 (1991)

\bibitem{makivic}
M. Makivi\'c and M. Jarrell, Phys. Rev. Lett. {\bf 68}, 1770 (1992).

\bibitem{qc} 
S.~Sachdev and J.~Ye, Phys. Rev. Lett. {\bf 69} 2411 (1992); 
A.~V. Chubukov and S.~Sachdev, Phys. Rev. Lett. {\bf 71}, 169 (1993);
A. V. Chubukov, S. Sachdev, and J. Ye, Phys. Rev. B {\bf 49}, 11919
(1994).

\bibitem{sokol}
A. Sokol, E. Gagliano, and S. Bacci, Phys. Rev. B {\bf 47}, 14646 (1993);
A. Sokol, R.L. Glenister, and R.R.P. Singh, Phys. Rev. Lett. {\bf 72}, 
1549 (1994).

\bibitem{qmcrates}
A. W. Sandvik and D. J. Scalapino, Phys. Rev. B {\bf 51}, 9403 (1995).

\bibitem{ramanexp1}
K. B. Lyons {\it et al.}, Phys. Rev. B {\bf 39}, 2293 (1989);
I. Ohana {\it et al.}, {\it ibid}. {\bf 39}, 2293 (1989);
P. E. Sulewski {\it et al.}, {\it ibid.} {\bf 41}, 225 (1990);
S. Sugai {\it et al.}, {\it ibid.}, {\bf 42}, 1045 (1990);

\bibitem{knoll}
P. Knoll {\it et al.}, Phys. Rev. B {\bf 42}, 4842 (1990). 

\bibitem{blumberg}
G. Blumberg {\it et al.}, Phys. Rev. B {\bf 53}, 11930 (1996).

\bibitem{parkinson}
J. B. Parkinson, J. Phys. C {\bf 2}, 2012 (1969).

\bibitem{canali}
C. M. Canali and S. M. Girvin, Phys. Rev. B {\bf 45}, 7127 (1992).

\bibitem{roger}
M. Roger, J. M. Delrieu, Phys. Rev. B. {\bf 39}, 2299 (1989).

\bibitem{diag}
S. Chakravarty, in {\it Proceedings of the Los Alamos Symposium 1989,
High Temperature Superconductivity}, edited by K. S. Bedell {\it et al.}
(Addison-Wesley, Reading MA, 1990); E. Gagliano and S. Bacci,
Phys. Rev. B {\bf 42}, 8772 (1990); E. Dagotto and D. Poilblanc,
{\it ibid.}, {\bf 42}, 7940 (1990); F. Nori, E. Gagliano, and S. Bacci,
Phys. Rev. Lett. {\bf 68}, 240 (1992).

\bibitem{fleury}
P. A. Fleury and R. Loudon, Phys. Rev. {\bf 166}, 514 (1968).

\bibitem{shastry}
B. S. Shastry and B. I. Shraiman, Phys. Rev. Lett. {\bf 65}, 1068 (1990);
Int. J. Mod. Phys. B {\bf 5}, 365 (1991).

\bibitem{singh}
R. R. P. Singh, P. A. Fleury, K. B. Lyons, and P. E. Sulewski,
Phys. Rev. Lett. {\bf 62}, 2736 (1989).

\bibitem{honda}
Y. Honda, Y. Kuramoto, and T. Watanabe, Physica C {\bf 185-189}, 1493
(1991); Phys. Rev. B {\bf 47}, 11329 (1993).

\bibitem{chubukov}
A. V. Chubukov and D. M. Frenkel, Phys. Rev. B {\bf 52}, 9760 (1995).

\bibitem{prelsingh}
R. R. P. Singh, P. Prelov\v{s}ek, and B. S. Shastry, Phys. Rev. Lett.
{\bf 77}, 4086 (1996).

\bibitem{gros}
C. Gros {\it et al.}, Phys. Rev. B {\bf 55}, 15048 (1996).

\bibitem{qmc1}
A.~W.~Sandvik and J.~Kurkij\"arvi, Phys. Rev. B {\bf 43}, 5950 (1991).

\bibitem{qmc2}
A.~W. Sandvik, J. Phys. A {\bf 25}, 3667 (1992).

\bibitem{sandvik}
A. W. Sandvik, Phys. Rev. B, to appear, Nov.~1 1997.

\bibitem{bacci}
S. Bacci and E. Gagliano, Phys. Rev. B {\bf 43}, 6224 (1991).

\bibitem{weber}
W. H. Weber and G. W. Ford, Phys. Rev. B {\bf 40}, 6890 (1989).

\bibitem{nori}
F. Nori {\it et al.}, Phys. Rev. Lett. {\bf 75}, 553 (1995).

\bibitem{manning}
S. Manning and F. V. Kusmartsev, J. Phys. Soc. Jpn {\bf 64}, 2245 (1995).

\bibitem{saenger}
D. U. Saenger, Phys. Rev. B {\bf 49}, 12176 (1994);
{\bf 52}, 1025 (1995).

\bibitem{dyson} 
F. J. Dyson, Phys. Rev. {\bf 102}, 1217 (1956); {\bf 102}, 1230 (1956).

\bibitem{maleev} 
S. V. Maleev, Zh. Eksp. Theor. Fiz. {\bf 30}, 1010 (1957)
[Sov. Phys. JETP {\bf 64}, 654 (1958)].

\bibitem{elliott} 
R. J. Elliott {\it et al.},  Phys. Rev. Lett. {\bf 21},
147 (1968).

\bibitem{oguchi} 
T. Oguchi, Phys. Rev. {\bf 117}, 117 (1960).

\bibitem{numz}
R. R. P. Singh, Phys. Rev. B {\bf 39}, 9760 (1989);
B. B. Beard and U.-J. Wiese, Phys. Rev. Lett. {\bf 77}, 5130 (1996).

\bibitem{maxent1} 
R. N. Silver, D. S. Sivia, and J. E. Gubernatis, Phys. Rev. B {\bf 41},
2380 (1990); J. E. Gubernatis, M. Jarrell, R. N. Silver, and
D. S. Sivia, {\it ibid.}, B {\bf 44} (1991) 6011.

\bibitem{maxent2} 
M. Jarrell and J. E. Gubernatis, Phys. Rep. {\bf 269}, 133 (1996).

\bibitem{schuttler}
H.-B. Sch\"uttler and D. J. Scalapino, Phys. Rev. Lett. {\bf 55},
1204 (1985); Phys. Rev. B {\bf 34}, 4744 (1986).

\bibitem{whitemoments}
S. R. White, Phys. Rev. B {\bf 44}, 4670 (1991).

\bibitem{bootstrap}
B. Efron and G. Gong, Am. Stat. {\bf 37}, 36 (1983).

\bibitem{previous}
In Ref.~\onlinecite{nori} some initial finite-temperature QMC results were
also presented. In these calculations the improved estimator introduced
here (see Appendix A) for $G(\tau)$ was not used, and the accuracy was 
therefore not as high as in the results presented here.

\bibitem{prelovsek}
P. Prelov\v{s}ek and J. Jakli\v{c}, Phys. Rev B {\bf 53}, 15095 (1996).

\bibitem{spin1}
P. A. Fleury and H. J. Guggenheim, Phys. Rev. Lett. {\bf 24}, 1346 (1970).

\bibitem{norinote}
The comparison with the experimental spectrum presented in Ref 
\protect{\onlinecite{nori}} 
(the same one, from Ref.~\protect{\onlinecite{singh}},
as used here for comparison in Fig.~\protect{\ref{figexperiment}})
is slightly affected by an accidental shift of the zero point of the 
frequency axis. With the correct zero point, the estimate for the 
strength of the disorder required decreases by $\approx 15-20$\%.

\bibitem{phononqmc}
A. W. Sandvik, R. R. P. Singh, and D. K. Campbell, to appear in
Phys. Rev. B (Nov.~1 1997).

\bibitem{rr}
M. J. Reilly and A. G. Rojo, Phys. Rev. B {\bf 53}, 6429 (1996).

\bibitem{handscomb}
D.~C.~Handscomb, Proc. Cambridge Philos. Soc. {\bf 58}, 594 (1962); 
{\bf 60}, 116 (1964).

\bibitem{lee}
D. H. Lee, J. D. Joannopoulos, and J. W. Negele, 
Phys. Rev. B {\bf 30}, 1599 (1984).

\bibitem{lnote}
In practice, it is useful to explicitly truncate the expansion at
some maximum $n=n_{\rm max}$, large enough to introduce no detectable 
systematical errors. One can then formulate a configuration space where 
the index sequence has a fixed length, by augmenting sequences shorter
than $n_{\rm max}$ by unit operators. This facilitates the construction
of a fast updating algorithm. See Refs.~\onlinecite{qmc1} and 
\onlinecite{sandvik}.

\bibitem{sizenote}
Scaling as $\beta N$, the index sequence is typically quite long at low
temperatures. For example, for a $10 \times 10$ lattice at inverse
temperature $\beta=80$, the average length $\langle n\rangle \approx 9500$. 
However, only one of the states $| \alpha (p)\rangle$ has to be stored.
The memory requirements are therefore significantly smaller than for
standard Trotter based methods.

\end{references}
\end{document}